\documentclass[10pt,conference]{IEEEtran}
\IEEEoverridecommandlockouts

\usepackage[ruled,vlined,linesnumbered]{algorithm2e}
\usepackage{enumitem}
\usepackage{subcaption}
\usepackage{booktabs} 
\usepackage{mychangebar}
\usepackage{enumitem}
\usepackage{xspace}
\usepackage{mdframed}
\usepackage{colortbl}
\usepackage{bigfoot}
\usepackage{verbatim}
\usepackage{hyperref}
\hypersetup{
  colorlinks=true,
  citecolor=black,
  filecolor=black,
  linkcolor=black,
  urlcolor=[cmyk]{1,0.58,0,0.21}  
}
\usepackage{cleveref}
\usepackage[final]{listings}
\usepackage{amssymb}
\usepackage{balance}

\usepackage{natbib-cite}
\usepackage{natbib-acm}

\SetArgSty{textnormal}

\makeatletter
\newcommand\footnoteref[1]{\protected@xdef\@thefnmark{\ref{#1}}\@footnotemark}
\makeatother

\usepackage{pgfplots}
\pgfplotsset{compat=1.13}
\usetikzlibrary{pgfplots.statistics}
\usepackage{pgf-pie}

\usepackage{todonotes}

\newcommand*{\MyIndent}{\hspace*{0.5cm}}%
\definecolor{light-gray}{gray}{0.85}
\definecolor{pythoncolor}{rgb}{1,.5,.5}
\definecolor{javacolor}{rgb}{.5,.5,1}

\newcommand{\mysec}[1]{\vspace{0.1cm} \noindent \textbf{#1.}}

\crefname{section}{Section}{Section}
\crefname{table}{Table}{Tables}
\crefname{figure}{Figure}{Figures}
\crefname{subfigure}{Figure}{Figures}
\crefname{definition}{Definition}{Definitions}
\crefname{equation}{Equation}{Equations}
\crefname{example}{Ex.}{Examples}
\crefname{algorithm}{Algorithm}{Algorithms}

\newcommand{\ie}[0]{i.e.,\xspace}
\newcommand{\eg}[0]{e.g.,\xspace}
\newcommand{\etc}[0]{etc.\xspace}

\newcommand{\tvistorrent}[0]{\textsc{TravisTorrent}\xspace}
\newcommand{\dockr}[0]{\textsc{Docker}\xspace}
\newcommand{\tvisyml}[0]{\texttt{{\small .travis.yml}}\xspace}
\newcommand{\bugswarm}[0]{\textsc{BugSwarm}\xspace}
\newcommand{\ghub}[0]{\textsc{GitHub}\xspace}
\newcommand{\git}[0]{\textsc{Git}\xspace}
\newcommand{\tvis}[0]{\textsc{Travis-CI}\xspace}
\newcommand{\tvisbuild}[0]{\texttt{{\small travis-build}}\xspace}
\newcommand{\pairminer}[0]{\textsc{Pair\-Miner}\xspace}
\newcommand{\pairfilter}[0]{\textsc{Pair\-Filter}\xspace}
\newcommand{\reproducer}[0]{\textsc{Repro\-ducer}\xspace}
\newcommand{\analyzer}[0]{\textsc{Ana\-lyzer}\xspace}

\newcommand{\textttsmall}[1]{\texttt{{\small #1}}\xspace}

\newcommand{\numJobPairArtifacts}{3,091\xspace}

\newcommand{\numProjects}{335\xspace}

\newcommand{\totalProjects}{670\xspace}

\newcommand{\mnlclasssize}[0]{\textsc{320}\xspace}
\newcommand{\nummnlclassctgrs}[0]{\textsc{10}\xspace}

\usepackage{enumitem} 
\setlist{leftmargin=5mm}
\setlist[itemize]{noitemsep}
\setlist[itemize]{nolistsep}

\begin{document}




\title{BugSwarm: Mining and Continuously Growing a Dataset of
  Reproducible Failures and Fixes} 


















\author{
\IEEEauthorblockN{
             David A. Tomassi\IEEEauthorrefmark{2}, Naji Dmeiri\IEEEauthorrefmark{2}, Yichen Wang\IEEEauthorrefmark{2}, Antara Bhowmick\IEEEauthorrefmark{2}
    }
    \IEEEauthorblockN{
Yen-Chuan Liu\IEEEauthorrefmark{2}, Premkumar T. Devanbu\IEEEauthorrefmark{2}, Bogdan Vasilescu\IEEEauthorrefmark{3}, Cindy Rubio-Gonz\'alez\IEEEauthorrefmark{2}
    }
    \IEEEauthorblockA{\IEEEauthorrefmark{2}University of California, Davis \{datomassi, nddmeiri, eycwang, abhowmick, yclliu, ptdevanbu, crubio\}@ucdavis.edu}
    \IEEEauthorblockA{\IEEEauthorrefmark{3}Carnegie Mellon University vasilescu@cmu.edu}

}


\maketitle

\begin{abstract}
Fault-detection, localization, and repair methods are vital to software quality;
but it is difficult to evaluate their generality, applicability, and current
effectiveness. Large, diverse, realistic datasets of durably-reproducible faults and fixes
are vital to good experimental evaluation of approaches to software quality, but
they are difficult and expensive to assemble and keep current. Modern continuous-integration (CI)
approaches, like \tvis, which are widely used, fully configurable, and executed
within custom-built containers, promise a path toward much larger defect
datasets. If we can identify and archive failing and subsequent passing runs,
the containers will provide a substantial assurance of durable future
reproducibility of build and test. Several obstacles, however, must be overcome
to make this a practical reality. We describe \bugswarm, a toolset that navigates these obstacles to
enable the creation of a \textit{scalable, diverse, realistic, continuously
growing} set of durably reproducible failing and passing versions of real-world,
open-source systems. The \bugswarm toolkit has already gathered
\numJobPairArtifacts fail-pass pairs, in Java and Python, all packaged within
fully reproducible containers. Furthermore, the toolkit can be run periodically
to detect fail-pass activities, thus growing the dataset continually. 
\end{abstract}



\section{Introduction}
\label{sec:intro}

Software defects have  major impacts on the economy, on
safety, and on the quality of life. Diagnosis and repair of software defects
consumes a great deal of time and money.
Defects can be treated more effectively, or avoided, by studying past
defects and their repairs. Several software engineering sub-fields, \eg
program analysis, testing, and automatic program repair, are dedicated to
developing tools, models, and methods for finding and repairing defects.
These approaches, ideally,  should be evaluated on realistic, up-to-date datasets of defects so
that potential users have an idea of how well they work. Such datasets should
contain \textit{fail-pass pairs}, consisting of a failing version,
which may include a test set that exposes the failure, and a passing version
including changes that repair it. Given this,
researchers can evaluate 
the effectiveness of tools that perform fault
detection, localization (static or dynamic), or fault repair. Thus,
research progress is intimately dependent on high-quality datasets of
fail-pass pairs.

There are several desirable properties of these datasets of fail-pass
pairs.  First, \textbf{scale}: enough data to attain
statistical significance on tool evaluations. Second,
\textbf{diversity}: enough variability in the data to control for
factors such as project scale, maturity, domain, language, defect
severity, age, \etc, while still retaining enough sample size for
sufficient experimental power. Third, \textbf{realism}: defects
reflecting actual fixes made by real-world programmers to repair real
mistakes. Fourth, \textbf{currency}: a continuously updated defect
dataset, keeping up with changes in languages, platforms, libraries,
software function, \etc, so that tools can be evaluated on bugs of
current interest and relevance. Finally, and most crucially, defect
data should be \textbf{durably reproducible}: defect data preserved in
a way that supports durable build and behavior reproduction,
robust to inevitable changes to libraries, languages, compilers,
related dependencies, and even the operating system.\footnote{While it
  is impossible to guarantee this in perpetuity, we would like to have
  some designed-in resistance to change.}

Some hand-curated datasets (\eg 
Siemens test suite~\citep{DBLP:conf/icse/HutchinsFGO94},
the SIR repository~\citep{DBLP:journals/ese/DoER05}, 
Defects4J~\citep{DBLP:conf/issta/JustJE14})
provide artifact collections to support controlled experimentation with program
analysis and testing techniques. However, these collections are curated by hand,
and are necessarily quite limited in \textit{scale} and \textit{diversity};
others incorporate small-sized student homeworks
~\citep{DBLP:journals/tse/GouesHSBDFW15}, which may not reflect development by
professionals. Some of these repositories often rely on seeded faults; natural
faults, from real programmers, would provide more \textit{realism}. At time of
creation, these are (or rather were) current. However, unless augmented through
continuous and expensive manual labor, \textit{currency} will erode. Finally, to
the extent that they have dependencies on particular versions of libraries and
operating systems, their future \textit{reproducibility} is uncertain.

The datasets cited above have incubated an impressive array of innovations and
are well-recognized for their contribution to research progress. However, we
believe that datasets of greater scale, diversity, realism, currency, and
durability will lead to even greater progress. The ability to control for 
covariates, without sacrificing experimental power, will help
tool-builders and empirical researchers obtain results with greater discernment,
external validity, and temporal stability. However, how can we build larger
defect datasets without heavy manual labor? Finding specific defect
occurrences, and creating recompilable and runnable versions of failing and
passing software is difficult for all but the most trivial systems: besides the
source code, one may also need to gather specific versions of libraries,
dependencies, operating systems, compilers, and other tools. This process requires a
great deal of human effort. Unless this human effort can somehow be automated
away, we cannot build large-scale, diverse, realistic datasets of reproducible
defects that continually maintain currency. But how can we automate this effort?

We believe that the DevOps- and OSS-led innovations in cloud-based
continuous integration (CI) hold the key. CI services, like \tvis
\citep{travis-ci}, allow
open-source projects to outsource integration testing.  OSS projects,
for various reasons, have need for continuous, automated integration
testing. In addition, modern practices such as test-driven development
have led to much greater abundance of automated tests. Every change to
a project can be intensively and automatically tested off-site, on a
cloud-based service; this can be done continually, across languages,
dependencies, and runtime platforms. For example, typical \ghub
projects require that each pull request (PR) be integration tested, and
failures fixed, before being vetted or merged by
integrators~\cite{gousios2015work, vasilescu2015quality}.  In active
projects, the resulting back-and-forth between PR contributors and
project maintainers naturally creates many fail-pass pair records in
the pull request history and overall project history.

Two key technologies underlie this capability: \textit{efficient, customizable,
container-based virtualization} simplifies handling of complex
dependencies, and \textit{scripted CI servers} allows custom automation of build
and test procedures. Project maintainers create scripts that define the test
environment (platforms, dependencies, \etc) for their projects; 
using these scripts, the cloud-based CI
services construct virtualized runtimes (typically \dockr containers) 
to build and run the
tests. The CI results are archived in ways amenable to mining and analysis. We
exploit precisely these CI archives, and the CI technology, to create an
automated, continuously growing, large-scale, diverse dataset of realistic and
durably reproducible defects.

In this paper, we present \bugswarm, a CI harvesting toolkit, together
with a large, growing dataset of durably reproducible defects.  The
toolkit enables maintaining currency and augmenting
diversity.
\bugswarm exploits archived CI log records to create detailed
artifacts, comprising buggy code versions, failing regression tests,
and bug fixes. When a successive pair of commits, the first, whose CI
log indicates a failed run, and the second, an immediately subsequent
passing run, is found, \bugswarm uses the project's CI customization
scripts to create an artifact: a fully containerized virtual
environment, comprising both versions and scripts to gather all
requisite tools, dependencies, platforms, OS, \etc \bugswarm artifacts
allow full build and test of pairs of failing/passing
runs. Containerization allows these artifacts to be durably
reproducible. The large scale and diversity of the projects using CI
services allows \bugswarm to also capture a large, growing, diverse,
and current collection of artifacts.

Specifically, we make the following contributions:
\begin{itemize}
\item We present an approach that leverages CI to mine fail-pass pairs
  in open source projects and automatically attempts to reproduce
  these pairs in \dockr containers (\cref{sec:infrastructure}).
\item We show that fail-pass pairs are frequently found in open source
  projects and discuss the challenges in reproducing such pairs
  (\cref{sec:evaluation}).
\item We provide the \bugswarm dataset of \numJobPairArtifacts
  artifacts, for Java and Python, to our knowledge the
  \textit{largest}, \textit{continuously expanding}, \textit{durably
    reproducible} dataset of fail-pass pairs, and describe the general
  characteristics of the \bugswarm artifacts
  (\cref{sec:evaluation}).\footnote{The \bugswarm dataset is available
    at \href{http://www.bugswarm.org}{http://www.bugswarm.org}.}
\end{itemize}

We provide background and further motivation for \bugswarm in
\cref{sec:background}. We describe limitations and future work in
\cref{sec:limitations}. Finally, we discuss related work in
\cref{sec:related} and conclude in \cref{sec:conclusions}.


\section{Background and Motivation}
\label{sec:background}

Modern OSS development, with CI services, provides an enabling
ecosystem of tools and data that support the creation of \bugswarm.
Here we describe the relevant components of this ecosystem and
present a motivating example.

\subsection{The Open-Source CI Ecosystem}

\mysec{\git and \ghub} \git is central to modern software
development. Each project has a \textit{repository}. Changes are added
via a \textit{commit}, which has a unique identifier, derived with a
SHA-1 hash. The project history is a sequence of commits. \git
supports branching.  The main development line is usually maintained
in a branch called \textttsmall{master}.  \ghub is a web-based service
hosting \git repositories.  \ghub offers \textit{forking}
capabilities, \ie cloning a repository but maintaining the copy
online.  \ghub supports the \textit{pull request} (PR) development
model: project maintainers decide on a case-by-case basis whether to
accept a change. Specifically, a potential contributor forks the
original project, makes changes, and then opens a pull request. The
maintainers review the PR (and may ask for additional changes) before
the request is merged or rejected.

\mysec{\tvis Continuous Integration} \tvis is the most popular
cloud-hosted CI service that integrates with \ghub; it can
automatically build and test commits or PRs.  \tvis is configured via
settings in a \tvisyml file in the project repository, specifying all
the environments in which the project should be tested.  A \tvis
\textit{build} can be initiated by a \textit{push event} or a
\textit{pull request event}. A push event occurs when changes are
pushed to a project's remote repository on a branch monitored by
\tvis. A pull request event occurs when a PR is opened and when
additional changes are committed to the PR.  \tvis builds run a
separate job for each configuration specified in the \tvisyml
file. The build is marked as ``passed'' when all its jobs pass.

\mysec{\dockr} \dockr is a lightweight virtual machine service that
provides application isolation, immutability, and customization. An
application can be packaged together with code, runtime, system tools,
libraries, and OS into an immutable, stand-alone, custom-built,
persistable \dockr \textit{image} (\textit{container}), which can be
run anytime, anywhere, on any platform that supports \dockr. In late
2014, \tvis began running builds and tests inside \dockr containers,
each customized for a specific run, as specified in the \tvis \tvisyml
files.  \tvis maintains some of its \textit{base} images containing a
minimal build environment.  \bugswarm harvests these containers to
create the dataset.

\subsection{Leveraging \tvis to Mine and Reproduce Bugs}
\label{sec:motivating}

\begin{figure}[t]
\center
  \includegraphics[scale=0.55]{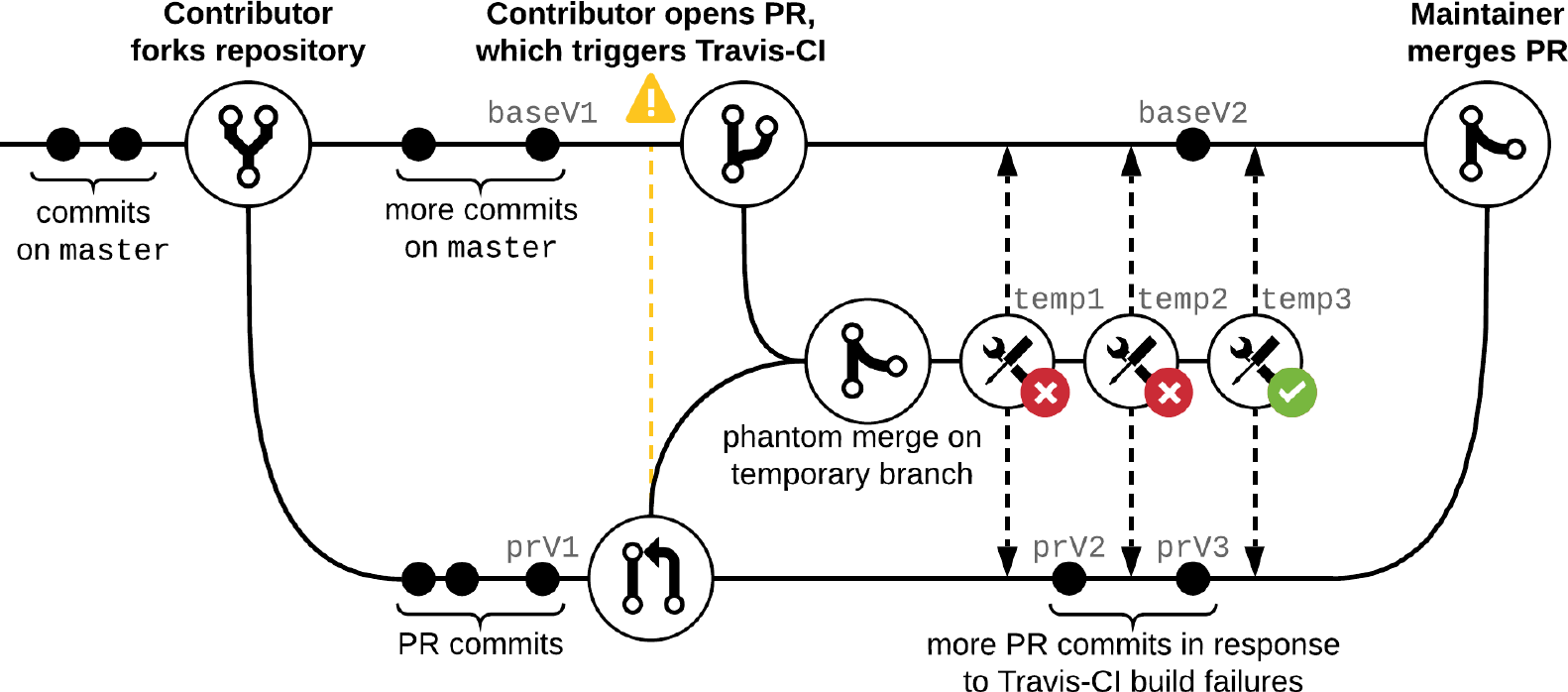}
  \caption{Lifecycle of a \tvis-built and tested PR}
  \label{fig:pr}
\end{figure}

We exploit \tvis to create \bugswarm. \cref{fig:pr} depicts the
lifecycle of a \tvis-built and tested PR. A contributor forks the
repository and adds three commits, up to \textttsmall{prV1}; she then
opens a PR, asking that her changes be merged into the original
repository. The creation of the PR triggers \tvis, which checks
whether there are merge conflicts between the PR branch and
\textttsmall{master} when the PR was opened (\textttsmall{prV1} and
\textttsmall{baseV1}). If not, \tvis creates a \textit{temporary}
branch from the base branch, into which the PR branch is merged to
yield \textttsmall{temp1}. This merge is also referred to as a
``phantom'' merge because it disappears from the \git history after
some time.\footnote{``Phantom'' merges present special challenges,
  which are discussed later.} \tvis then generates build scripts from
the \tvisyml file and initiates a build, \ie runs the scripts to
compile, build, and test the project.

In our example, test failures cause the first build to fail; \tvis
notifies the contributor and project maintainers, as represented by
the dashed arrows in \cref{fig:pr}. The contributor does her fix and
updates the PR with a new commit, which triggers a new build. Again,
\tvis creates the merge between the PR branch (now at
\textttsmall{prV2}) and the base branch (still at
\textttsmall{baseV1}) to yield \textttsmall{temp2}. The build fails
again; apparently the fix was no good. Consequently, the contributor
updates the PR by adding a new commit, \textttsmall{prV3}. A \tvis
build is triggered in which the merge (\textttsmall{temp3}) between
the PR branch (at \textttsmall{prV3}) and the base branch (now at
\textttsmall{baseV2}) is tested.\footnote{\tvis creates each phantom
  merge on a separate temporary branch, but \cref{fig:pr} shows the
  phantom merges on a single branch for simplicity.}  This time, the build
passes, and the PR is accepted and merged into the base branch.

Each commit is recorded in version control, archiving source
code at build-time plus the full build
configuration (\tvisyml file). \tvis records how each build fared (pass or
fail) and archives a build log containing output of the build and test process,
including the names of any failing tests.

Our core idea is that \tvis-built and tested pull requests (and
regular commits) from \ghub, available in large volumes for a variety
of languages and platforms, can be used to construct \textit{fail-pass
  pairs}. In our example, the version of the code represented by the
merge \textttsmall{temp2} is ``defective,'' as documented by test
failures in the corresponding \tvis build log. The subsequently
``fixed'' version (no test failures in the build log) is represented by
\textttsmall{temp3}. Therefore, we can extract (1)~a failing program
version; (2)~a subsequent, fixed program version; (3)~the fix, \ie the
difference between the two versions; (4)~the names of failing tests
from the failed build log; (5)~a full description of the build
configuration.

Since each \tvis job occurs within a \dockr container, we can
re-capture that specific container image, thus rendering the
event durably reproducible. Furthermore, if one could build an \textit{automated
harvesting system} that could continually mine \tvis builds and create \dockr
images that could persist these failures and fixes, this promises a way to
create a dataset to provide all of our desired data: \ghub-level \textit{scale};
\ghub-level \textit{diversity}; \textit{realism} of popular OSS projects;
\textit{currency} via the ability to automatically and periodically augment our
dataset with recent events, and finally \textit{durable reproducibility} via
\dockr images.


\section{\bugswarm Infrastructure}
\label{sec:infrastructure}

\subsection{Some Terminology}
A project's \textit{build history} refers to all \tvis builds
previously triggered.  A \textit{build} may include many
\textit{jobs}; for example, a build for a Python project might include
separate jobs to test with Python versions 2.6, 2.7, 3.0, \etc

A \textit{commit pair} is a 2-tuple of \git commit SHAs that each
triggered a \tvis build in the \textit{same} build history. The
canonical commit pair consists of a commit whose build fails the tests
followed by a fix commit whose build passes the tests.  The terms \textit{build
  pair} and \textit{job pair} refer to a 2-tuple of \tvis builds or
jobs, respectively, from a project's build history. For a given build,
the \textit{trigger commit} is the commit that, when pushed to the
remote repository, caused \tvis to start a build.

\bugswarm has four components: \pairminer, \pairfilter, \reproducer,
and \analyzer. These components form the pipeline that curates \bugswarm
artifacts and are designed to be relatively independent and general.
This section describes the responsibilities and implementation of each
component, and a set of supporting tools that facilitate usage of the
dataset.

\subsection{Design Challenges}
The tooling infrastructure is designed to handle certain specific
challenges, listed below, that arise when one seeks to continuously
and automatically mine \tvis. In each case, we list the tools that
actually address the challenges.

\mysec{Pair coherence} Consecutive commits in a \git history may not
correspond to consecutive \tvis builds. A build history, which \tvis
retains as a linear series of builds, must be traversed and transformed
into a directed graph so that pairs of consecutive builds map
to pairs of consecutive commits.  \git's non-linear nature makes this
non-trivial. (\pairminer)

\mysec{Commit recovery} To reproduce a build, one needs to find the
trigger commit. There are several (sub-)challenges here. First,
temporary merge commits like \texttt{temp1,2,3} in \cref{fig:pr} are
the ones we need to extract, but these are not retained by \tvis.
Second, \git's powerful history-rewriting capabilities allow commits
to be erased from history; developers can and do collapse commits like
\texttt{prV1,2,3} into a single commit, thus frustrating the ability
to recover the consequent phantom merge commits.  (\pairminer,
\pairfilter)

\mysec{Image recovery}
In principle, \tvis creates and retains \dockr images that allow re-creation of
build and test events. In practice, these images are not always archived as
expected and so must be reconstructed. (\pairfilter, \reproducer)

\mysec{Runtime recovery}
Building a specific project version often requires satisfying a large
number of software dependencies on tools, libraries, and frameworks; all or some
of these may have to be ``time-traveled'' to an earlier version. (\reproducer)

\mysec{Test flakiness}
Even though \tvis test behavior is theoretically recoverable via \dockr images, tests may
behave non-deterministically because of concurrency or environmental
(\eg external web service) changes. Such flaky tests lead to flaky builds, which both
must be identified for appropriate use in experiments. (\reproducer)

\mysec{Log analysis}
Once a pair is recoverable, \bugswarm tries to determine the exact
nature of the failure from the logs, which are not well structured and have
different formats for each language, build system, and test toolset combination.
Thus the logs must be carefully analyzed to recover the nature of the failure
and related metadata (\eg raised exceptions, failed test names, \etc), so
that the pair can be documented. (\analyzer)


\begin{algorithm}[t]
\small
  \KwIn{Project slug $P$}
  \KwOut{Set $J$ of fail-pass job pairs $(j_f, j_p)$}

    $J$ = $\emptyset$\, $B$ = the list of \tvis builds for $P$\;
    G = \{g $\mid$ g $\subseteq B$ and $\forall b \in g$ belong to the same branch/PR\}\;
    \ForEach{$g$ in $G$}{
        Order the builds in $g$ chronologically\;
        \ForEach{$b_i \in g$}{
            \If{$b_i$ is failed and $b_{i+1}$ is passed}{
                AssignCommits($b_i$)\;
                AssignCommits($b_{i+1}$)\;
                $J$ = $J \cup \{(j_f, j_p)$ $\mid$ $j_f \in b_i$ and $j_p \in b_{i+1}$ and $j_f$ has the same configuration as $j_p$\}\;
            }
        }
    }
    \Return{$J$}
  \caption{\pairminer Algorithm}
  \label{alg:pair-miner}
\end{algorithm}

\subsection{Mining Fail-Pass Pairs}
\label{sec:pair-miner}

\pairminer extracts from a project's \git and build histories a set of
fail-pass job pairs (\cref{alg:pair-miner}). \pairminer takes as
input a \ghub slug and produces a set of fail-pass job pairs annotated
with trigger commit information for each job's parent build. The \pairminer
algorithm involves (1)~delinearizing the project's build history, (2)~extracting
fail-pass build pairs, (3)~assigning commits to each pair, and (4)~extracting
fail-pass job pairs from each fail-pass build pair.

\mysec{Analyzing build history}
\pairminer first downloads the project's entire build history with the \tvis
API. For each build therein, \pairminer notes the branch and (if
applicable) the pull request containing the trigger commit,
\pairminer first resolves the build history into lists of builds that were triggered
by commits on the same branch or pull request. 
\pairminer recovers the
results of the build and its jobs (passed, failed, errored, or canceled), the
\tvisyml configuration of each job, and the unique identifiers of the build and
its jobs using the \tvis API.

\mysec{Identifying fail-pass build pairs}
Using the build and job identifiers, \pairminer finds consecutive pairs
where the first build failed and the second passed. Builds are
considered from all branches, including the main line and any perennials, and
both merged and unmerged pull requests. Next, the triggering commits
are found, and recovered from \git history.

\begin{algorithm}[t]
\small
  \KwIn{\tvis build $B$}

    Mark $B$ as ``unavailable'' by default\;
    Clone the \git repository for B\;
    \If{$B$ is triggered by a push event}{
        Assign trigger commit $t$ from \tvis build metadata\;
        \If{$t$ in \git history or $t$ in \ghub archive}{
            mark $B$ as ``available''\;
        }
    }
    \ElseIf{$B$ is triggered by a pull request event}{
        Assign trigger commit $t$, base commit $b$, and merge commit $m$ for $B$ from \tvis build metadata\;
        \If{$t$ and $b$ in \git history or $m$ in \ghub archive}{
            mark $B$ as ``available''\;
        }
    }
  \caption{AssignCommits Algorithm}
  \label{alg:assigncommits}
\end{algorithm}

\mysec{Finding trigger commits} If the trigger commit was a push
event, then \pairminer can find its SHA via the \tvis API.  For pull
request triggers, we need to get the pull request and base branch head
SHAs, and re-create the phantom merge. Unfortunately, neither the
trigger commit nor the base commit are stored by \tvis; recreating
them is quite a challenge. Fortunately, the commit message of the
phantom commit, which is stored by \tvis, contains this information; we
follow \citet{msr17challenge} to extract this information.  This
approach is incomplete but is the best available.

\tvis creates temporary merges for pull-request builds. While
temporary merges may no longer be directly accessible, the information
for such builds (the head SHAs and base SHAs of the merges) are
accessible through the \ghub API. We resort to \ghub archives to
retrieve the code for the commits that are no longer in \git history.

Even if the trigger commit is recovered from the phantom merge commit,
one problem remains: developers might squash together all commits in a
pull request, thus erasing the constituent commits of the phantom
merge right out of the \git history. In addition, trigger commits for
push event builds can sometimes also be removed from the \git history
by the project personnel.  As a result, recreating this merge is not
always possible; we later show the proportion for which we were able
to reset the repository to the commits in the fail-pass pairs. The two
steps of phantom recovery\textemdash first finding the trigger commits and then
ensuring that the versions are available in \ghub\textemdash are described
in~\cref{alg:assigncommits}.

\mysec{Extracting fail-pass job pairs} \pairminer now has a list of
fail-pass build pairs for the project. As described in
\cref{sec:background}, each build can have many jobs, one for each
supported environment.  A build fails if any one of its jobs fails and
passes if all of its jobs pass.  Given a failing build, \pairminer
finds pairs of jobs, \textit{executed in the same environment}, where
the first failed and the second passed. Such a pair only occurs when
a defective version was fixed via source code patches and not by
changes in the execution environment (see~\cref{alg:pair-miner}).


\subsection{Finding Essential Information for Reproduction}
\label{sec:pair-filter}

Pairs identified by \pairminer must be assembled into reproducible
containers.  To stand a chance of reproducing a job, one must have
access to, at a minimum, these essentials: (1)~the state of the
project at the time the job was executed and (2)~the environment in
which the job was executed. For each job in the pipeline, \pairfilter
checks that these essentials can be obtained.  If the project state
was deemed recoverable by \pairminer, \pairfilter retrieves the
original \tvis log of the job and extracts information about the
execution environment. Using timestamps and instance names in the log,
\pairfilter determines if the job executed in a \dockr container and,
if so, whether the corresponding image is still accessible. If the log
is unavailable, the job was run before \tvis started using \dockr, or
the particular image is no longer publicly accessible, then the job is
removed from the pipeline.


\subsection{Reproducing Fail-Pass Pairs}
\label{sec:reproducer}
\reproducer checks if each job is durably reproducible.
This takes several steps, described below.

\mysec{Generating the job script}
\tvisbuild,\footnote{\href{https://github.com/travis-ci/travis-build}{https://github.com/travis-ci/travis-build}}
a component of \tvis, produces a shell script from a \tvisyml file for
running a \tvis job.  \reproducer then alters the script to reference
a specific past version of the project, rather than the latest.

\mysec{Matching the environment}
\label{sec:reconstructing-environment}
To match the original job's runtime environment, \reproducer chooses
from the set of \tvis's publicly available \dockr images, from Quay
and DockerHub, based on (1)~the language of the project, as indicated
by its \tvisyml configuration, and (2)~a timestamp and instance name
in the original job log that indicate when that image was built with
\dockr's tools.

\mysec{Reverting the project} For project history reversion,
\reproducer clones the project and resets its state using the trigger
commit mined by \pairminer. If the trigger was on a pull request,
\reproducer re-creates the phantom merge commit using the trigger and
base commits mined by \pairminer.  If any necessary commits were not
found during the mining process, \reproducer downloads the desired
state of the project directly from a zip archive maintained by
\ghub.\footnote{ \ghub allows one to download a zip
  archive of the entire project's file structure at a
  specific commit. Since this approach produces a stand-alone checkout
  of a project's history (without any of the \git data
  stores), \reproducer uses this archive only if a proper clone and
  reset is not possible.} Finally, \reproducer plants the state of the
project inside the execution environment to reproduce the job.

\mysec{Reproducing the job} \reproducer creates a new \dockr image, as
described in~\cref{sec:reconstructing-environment}, runs the generated
job script, and saves the resulting output stream in a log
file. \reproducer can run multiple jobs in parallel. \reproducer
collects the output logs from all the jobs it attempts to reproduce
and sends them to \analyzer for parsing.

\begin{table}[t]
\small
  \caption{\bugswarm's main metadata attributes}
  \label{tab:attributes}
  \centering
  \setlength{\tabcolsep}{4pt}
  \begin{tabular}{p{2cm}p{6cm}}
    \toprule

    Attribute Type & Description \\
    \midrule
    Project &  \ghub slug, primary language, build system, and
              test framework\\
    Reproducibility &  Total number of attempts, and number
                      of successful attempts to reproduce pair\\
    Pull Request &  Pull request \#, merge timestamp, and branch\\
    \tvis Job &  \tvis build ID, \tvis job ID, number of executed
                 and failed tests, names of the failed tests, trigger commit, and branch name\\
    Image Tag &  Unique \textit{image tag} (simultaneously serves as a reference to a
                particular \dockr image)\\
    \bottomrule
  \end{tabular}
\end{table}

\begin{table*}[t]
\small
  \caption{Mined Fail-Pass Pairs}
  \label{tab:mined-pairs}
  \centering
  \setlength{\tabcolsep}{4pt}
  \begin{tabular}{ lrrrrrrrrrrrrr }
    \toprule

    & \multicolumn{5}{c}{Push Events} && \multicolumn{5}{c}{Pull Request Events} \\
    \cmidrule{2-6}\cmidrule{8-12}

    Language    & Failed Jobs & All Pairs & Available & \dockr & w/Image   &  & Failed Jobs & All Pairs & Available & \dockr & w/Image   \\
\midrule
Java        &     320,918 &    80,804 &    71,036 &  50,885 & 29,817 &  &     250,349 &    63,167 &    24,877 &  20,407 &  9,509 \\
Python       &     778,738 &   115,084 &   103,175 &  65,924 & 37,199 &  &   1,190,186 &   188,735 &    62,545 &  45,878 & 24,740 \\
\midrule
Grand Total  &   1,099,656 &   195,888 &   174,211 & 116,809 & \textbf{67,016} &  &   1,440,535 &   251,902 &    87,422 &  66,285 & \textbf{34,249} \\

    \bottomrule
  \end{tabular}
\end{table*}


\subsection{Analyzing Results}
\label{sec:analyzer}

\analyzer parses a \tvis build log to learn the status of the build
(passed, failed, \etc) and the result of running the regression test
suite. If there are failing tests, then \analyzer also retrieves their
names. A challenge here: the format of build logs varies substantially
with the specific build system and the test framework; so parsers must
be specialized to each build and test framework.  For Java, we support
the most popular build systems---Maven \citep{maven}, Gradle
\citep{gradle}, and Ant \citep{ant}---and test frameworks---JUnit
\citep{junit} and testng \citep{testng}. For Python, we support the
most popular test frameworks---unittest \citep{unittest}, unittest2
\citep{unittest2}, nose \citep{nose}, and pytest \citep{pytest}.

\analyzer has a top-level analyzer that retrieves all
language-agnostic items, such as the operating system used for a
build, and then delegates further log parsing to
language-specific and build system-specific analyzers that extract
information related to running the regression test suite. The extracted
attributes\textemdash number of tests passed, failed, and skipped; names
of the failed tests (if any); build system, and test framework\textemdash
are used to compare the original
\tvis log and the reproduced log. If the attributes match, then we say
the run is reproducible. Writing a new language-specific analyzer is
relatively easy, mostly consisting of regular expressions that capture
the output format of various test frameworks.

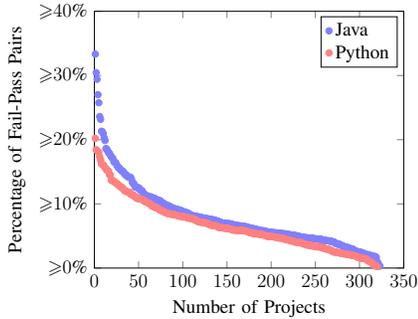
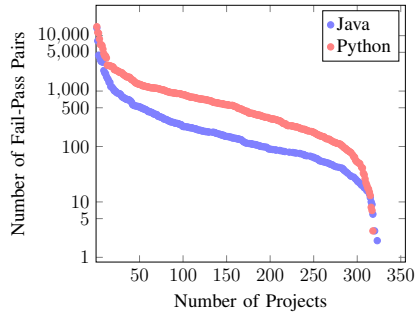
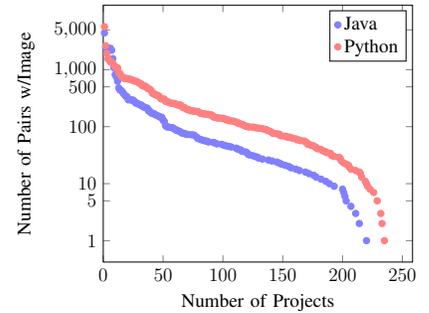
\begin{figure*}
\begin{subfigure}[b]{0.33\textwidth}
	\begin{center}
		\begin{tikzpicture}[scale=0.6]
            \large
			\begin{axis}[
				legend cell align={left},
				ylabel = Percentage of Fail-Pass Pairs,
				y label style={at={(1, 0)}},
				xlabel = Number of Projects,
				x label style={at={(0, 1)}},
                                ylabel near ticks,
                                xlabel near ticks,
				xmin=0,
				ymax=40,
				ymin=0,
				xmax=350,
				yticklabel={$\geqslant$\pgfmathprintnumber\tick\%},
				]
		\legend{Java, Python}
		\addplot[
		scatter,
		only marks,
		point meta=explicit symbolic,
		scatter/classes={
		j={mark=*,javacolor},%
		p={mark=*,pythoncolor}},
		]
			table[x=num_pairs, y=percent, meta=label] {figures/data/fail-pass-percent-per-language-cumulative.tsv};
		\end{axis}
		\end{tikzpicture}
	\end{center}
	\caption{Percent of Fail-Pass Pairs per Language}
	\label{fig:miner-percent}
\end{subfigure}
\begin{subfigure}[b]{0.33\textwidth}
	\begin{center}
		\begin{tikzpicture}[scale=0.6]
		    \large
		    \begin{axis}[
		    	legend cell align={left},
		        xlabel=Number of Projects,
		        x label style={at={(1, 0)}},
		        ylabel=Number of Fail-Pass Pairs,
		        y label style={at={(axis description cs:-0.21,.5)},rotate=0,anchor=south},
                                xlabel near ticks,
		        ymode=log,
		        log ticks with fixed point,
		        xmin=0.1,
				ymin=0,
				ytick={0,1,5,10,100, 500, 1000, 5000, 10000},
				]
		\legend{Java, Python}
        \addplot[
        scatter,
		only marks,
		point meta=explicit symbolic,
		scatter/classes={
		j={mark=*,javacolor},
		p={mark=*,pythoncolor}},
		]
        table[y=num_pairs, x=num_projects, meta=label] {figures/data/fail-pass-number-per-language-cumulative.tsv};
		    \end{axis}
		\end{tikzpicture}
	\end{center}
	\caption{Cumulative Number of Fail-Pass Pairs}
	\label{fig:miner-number}
\end{subfigure}
\begin{subfigure}[b]{0.33\textwidth}
	\begin{center}
		\begin{tikzpicture}[scale=0.6]
		    \large
		    \begin{axis}[
		    	legend cell align={left},
		        xlabel=Number of Projects,
		        x label style={at={(1, 0)}},
		        ylabel=Number of Pairs w/Image,
		        y label style={at={(axis description cs:-0.2,.5)},rotate=0,anchor=south},
                                xlabel near ticks,
		        ymode=log,
		        log ticks with fixed point,
		        xmin=0,
				ymin=0,
				ytick={0,1,5,10,100, 500, 1000, 5000},
				]
		\legend{Java, Python}
        \addplot[
        scatter,
		only marks,
		point meta=explicit symbolic,
		scatter/classes={
		j={mark=*,javacolor},
		p={mark=*,pythoncolor}},
		]
        table[y=num_pairs, x=num_projects, meta=label] {figures/data/unfiltered-number-per-language-cumulative.tsv};
		    \end{axis}
		\end{tikzpicture}
	\end{center}
	\caption{Cumulative Number of Pairs w/ Image}
	\label{fig:unfiltered-number}
\end{subfigure}
\caption{Frequency of Fail-Pass Pairs}
\label{fig:miner}
\end{figure*}

\subsection{Tools for \bugswarm Users}
\bugswarm includes tools to support tasks such as artifact selection,
artifact retrieval, and artifact execution.

\mysec{Artifact selection \& retrieval} A given experiment may require
artifacts meeting specific criteria.  For this reason, each artifact
includes metadata as described in~\cref{tab:attributes}.  The
\bugswarm website provides an at-a-glance view of the metadata for all
artifacts.  Simple filtering can be done directly via the web
interface. For more advanced filtering, we provide a REST API; a
Python API is also available.

To facilitate retrieval of artifact \dockr images, we provide a
\bugswarm command line interface that masks the complexities of the
\dockr ecosystem to use our artifacts.  Given any \bugswarm artifact
identifier, the CLI can download the artifact image, start an
interactive shell inside the container, and clean up the container
after
use.\footnote{\href{https://github.com/BugSwarm/client}{https://github.com/BugSwarm/client}}

\mysec{Artifact execution}
\label{sec:artifact-execution}
A typical workflow for experiments with \bugswarm involves copying
tools and scripts into a container, running jobs, and then copying
results. We provide a framework to support this common artifact
processing workflow. The framework can be extended to fit users'
specific needs. See the \bugswarm website for example applications.


\section{Experimental Evaluation}
\label{sec:evaluation}

Our evaluation is designed to explore the feasibility of automatically
creating a large-scale dataset of reproducible bugs and their
corresponding fixes. In particular, we answer the following research
questions:
\begin{enumerate}[label=\textbf{RQ\arabic*:},wide,labelwidth=!,labelindent=0pt]
\item How often are fail-pass pairs found in OSS projects?
\item What are the challenges in automatically reproducing fail-pass pairs?
\item What are the characteristics of reproducible pairs?
\end{enumerate}

The \bugswarm infrastructure is implemented in Python. \reproducer
uses a modified version of the \tvisbuild component from \tvis to
translate \tvisyml files into shell scripts. The initial Java-specific
\analyzer was ported to Python from
\tvistorrent's~\citep{msr17challenge} implementation in
Ruby. \analyzer has been extended to support JUnit for Java and now
also supports Python.

\bugswarm requires that a project is hosted on \ghub and uses
\tvis. We randomly selected \numProjects projects among the 500 \ghub
projects with the most \tvis builds, for each of Java and Python.

\begin{table*}[t]
\small
  \caption{Reproduced Pairs}
  \label{tab:reproduced-pairs}
  \centering
  \setlength{\tabcolsep}{4pt}
  \begin{tabular}{ lrrrrrrrrrr }
    \toprule

    && \multicolumn{4}{c}{Fully Reproducible + Flaky} &&&& \\
    \cmidrule{3-6}

    Language    & Pairs to Reproduce &  w/Failed Test & w/Failed Job &
Error-Pass & Total Pairs &  & Unreproducible & Pending\\
\midrule
Java        &             39,326 &                 584 + 15 & 564 + 22&        626 + 16 & 1,827 &  &          17,369 & 20,130 \\
Python      &             61,939 &                 785 + 41 & 387 +\hspace{0.5em} 3&48 +\hspace{0.5em} 0 & 1,264 &  &         35,126& 25,549 \\
\midrule
Grand Total &            101,265 &               1,425 &          976&        690 & \textbf{3,091} &  &          52,495& 45,679 \\

    \bottomrule
  \end{tabular}
\end{table*}

\pgfplotstableread[row sep=\\,col sep=&]{
    type            & java  & python  \\
    w/Failed Test &   599 &   826 \\
    w/Failed Job    &   586 &    390 \\
    Error-Pass      &   642 &    48 \\
}\bargraphdata
\begin{figure*}
\begin{subfigure}[b]{0.33\textwidth}
	\begin{center}
		\begin{tikzpicture}[scale=0.6]
            \large
			\begin{axis}[
				legend cell align={left},
				ylabel = Percentage of Reproduced Pairs,
				y label style={{at={(axis description cs:-.21,.5)},rotate=0,anchor=south}},
				xlabel = Number of Projects,
				x label style={at={(0, 1)}},
                                xlabel near ticks,
				xmin=0,
				ymax=100,
				ymin=-1,
				xmax=120,
				ytick={0,20, 40, 60, 80,100},
				xtick={20, 40, 60, 80,100, 120},
				yticklabels={$\geqslant$0\%, $\geqslant$20\%, $\geqslant$40\%, $\geqslant$60\%, $\geqslant$80\%, 100\%}
				]
		\legend{Java, Python}
		\addplot[
		scatter,
		only marks,
		point meta=explicit symbolic,
		scatter/classes={
		j={mark=*,javacolor},%
		p={mark=*,pythoncolor}},
		]
		table[x=num_pairs, y=percent, meta=label] {figures/data/reproduced-percent-per-language-cumulative.tsv};
		\end{axis}
		\end{tikzpicture}
	\end{center}
	\caption{Cumulative Percentage of Reprod. Pairs}
	\label{fig:reproducer-percent}
\end{subfigure}
\begin{subfigure}[b]{0.33\textwidth}
	\begin{center}
		\begin{tikzpicture}[scale=0.6]
		    \large
		    \begin{axis}[
		    	legend cell align={left},
		        xlabel=Number of Projects,
		        x label style={at={(1, 0)}},
		        ylabel=Number of Reproduced Pairs,
		        y label style={at={(axis description cs:-0.14,.5)},rotate=0,anchor=south},
                xlabel near ticks,
		        ymode=log,
		        log ticks with fixed point,
		        xmin=0.1,
				ymin=0,
				ytick={0,1,5,10, 50, 100, 500},
				]
		\legend{Java, Python}
        \addplot[
        scatter,
		only marks,
		point meta=explicit symbolic,
		scatter/classes={
		j={mark=*,javacolor},
		p={mark=*,pythoncolor}},
		]
        table[y=num_pairs, x=num_projects, meta=label] {figures/data/reproduced-number-per-language-cumulative.tsv};
		    \end{axis}
		\end{tikzpicture}
	\end{center}
	\caption{Cumulative Number of Reprod. Pairs}
	\label{fig:reproducer-number}
\end{subfigure}
\begin{subfigure}[b]{0.33\textwidth}
    \begin{center}
        \begin{tikzpicture}[scale=0.6]
            \large
            \begin{axis}[
                    legend cell align={left},
                    ybar,
                    bar width=.6cm,
                    enlarge x limits={abs=1.2cm},
                    legend style={at={(0.84,.97)},
                        anchor=north,legend columns=1},
                    symbolic x coords={w/Failed Test,w/Failed Job,Error-Pass},
                    xtick=data,
                    x tick label style={ anchor=north, align=center,text width=2cm},
                    nodes near coords,
                    nodes near coords align={vertical},
                    ymin=0,
                    ymax=1000,
                    ylabel={Number of Pairs},
                    ytick={0, 200, 400, 600, 800},
                    y label style={at={(axis description cs:-0.155,.5)},anchor=south},
                ]
                \addplot+[javacolor,fill,text=black] table[x=type,y=java]{\bargraphdata};
                \addplot+[pythoncolor,fill,text=black] table[x=type,y=python]{\bargraphdata};
                \legend{Java, Python}
            \end{axis}
        \end{tikzpicture}
	\end{center}
	\caption{Breakdown of Reproduced Pairs}
	\label{fig:reproducer-types}
\end{subfigure}
\caption{Reproduced Pairs}
\label{fig:reproducer}
\end{figure*}
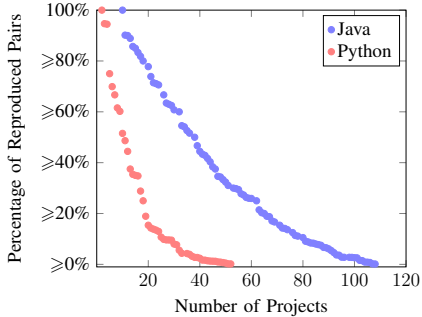
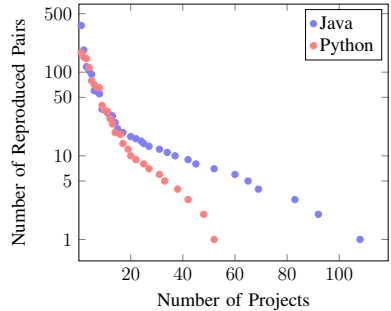
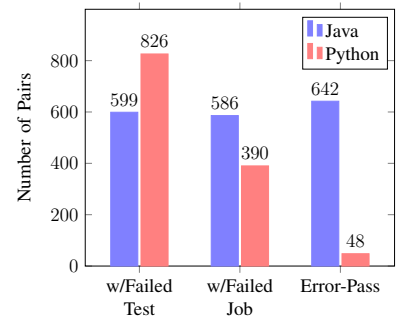

\subsection{Mining Fail-Pass Pairs}

We inspected a total of 10,179,558 jobs across \totalProjects
projects, from which 2,540,191 are failed jobs. We mined a total of
447,790 fail-pass pairs. As described in \cref{sec:pair-miner}, pairs
can originate from push events or pull request
events. \cref{tab:mined-pairs} shows the breakdown: push events
contribute to 44\% of the fail-pass pairs (195,888) and pull requests
contribute to 56\% (251,902). Note that fail-pass pairs represent an
under-approximation of the number of bug-fix commits; \bugswarm pairs
do not capture commits that fix a bug whose build is not broken. We
calculate the percentage of fail-pass pairs with respect to the total
number of successful jobs (potential fixes to a bug) per
project. \cref{fig:miner-percent} plots a cumulative graph with the
results. In general, we find that Java projects have a slightly higher
percentage of fail-pass pairs (at most 33\%) than Python projects (at
most 20\%). For example, there are 80 Java projects and 61 Python
projects for which at least 10\% of the passing jobs fix a
build. \cref{fig:miner-number} plots the cumulative \textit{number} of
fail-pass pairs per project. The Java and Python projects with the
most pairs have 13,699 and 14,510 pairs, respectively.

We run \pairfilter to discard fail-pass pairs that are unlikely to be
reproducible. \cref{tab:mined-pairs} shows the number of
fail-pass pairs after each filter is applied. Specifically, columns
``Available'' show the pairs we can reset to or which are archived,
columns ``\dockr'' show the number of remaining pairs that use
a \dockr image, and columns ``w/Image'' show the number of remaining
pairs for which we can locate \tvis base
images. \cref{fig:unfiltered-number} plots the cumulative
\textit{number} of w/Image pairs, which are passed to \reproducer. A
total of 220 Java projects and 233 Python projects have w/Image pairs.

\begin{mdframed}[backgroundcolor=gray!20]
  \small \textbf{RQ1}: At most 33\% and 22\% of all pairs of Java and
  Python projects, respectively, follow the fail-pass pattern
  (\cref{fig:miner}). Among 670 projects, we find a total of 447,490
  fail-pass pairs, from which 101,265 pairs may be reproducible.
\end{mdframed}

\subsection{Reproducing Fail-Pass Pairs}

We successfully reproduced \numJobPairArtifacts out of 55,586
attempted pairs (45,679 pairs are pending reproduction due to time
constraints). Recall from \cref{sec:pair-miner} that \pairminer mines
\textit{job pairs}. The corresponding number of reproducible unique
\textit{build pairs} is 1,837 (1,061 for Java and 776 for Python). The
rest of the paper describes the results in terms of number of job
pairs. The \numJobPairArtifacts artifacts belong to 108 Java projects
and 52 Python projects. \cref{tab:top-projects} lists the 5 projects
with the most artifacts for each language. We repeated the
reproduction process 5 times for each pair to determine its
stability. If the pair is reproducible all 5 times, then it is marked
as ``reproducible.'' If the pair is reproduced only sometimes, then it
is marked as ``flaky.''  Otherwise, the pair is said to be
``unreproducible.'' Numbers for each of these categories can be found
in \cref{tab:reproduced-pairs}.

\cref{fig:reproducer-percent} shows the cumulative percentage of
reproduced pairs across projects. We achieve a 100\% pair reproduction
rate for 10 Java projects and 2 Python projects, at least 50\% for 38
Java projects and 50 Python projects, and at least 1 pair is
reproducible in 108 Java projects, and 52 Python
projects. \cref{fig:reproducer-number} shows the cumulative
\textit{number} of reproduced pairs. The Java and Python projects with
the most reproducible pairs have 361 and 171, respectively.

We further classify ``reproducible'' and ``flaky'' pairs into three
groups: (1)~pairs that have failed tests, (2)~pairs that do not have
failed tests despite a failed build, and (3)~pairs whose build
finishes with an error. (1)~and (2)~are labeled \textit{failed} and
(3)~\textit{errored}. This naming convention is from \tvis
\citep{travis-ci-build-result} and is defined by the part of the job
lifecycle that encounters a non-zero exit code.  Typically,
\textit{errored} builds have dependency-related issues.
\cref{fig:reproducer-types} shows the breakdown for both Java and
Python. We find that 46.1\%, 31.6\%, and 22.3\% of reproducible pairs
correspond to each of the above categories, respectively.

Surprisingly, only 97 pairs were ``flaky.'' We suspect a number of
unreproducible pairs are indeed flaky but running them 5 times was not
sufficient to identify them. We plan to investigate how to grow the
number of flaky pairs in \bugswarm. An initial direction could involve
selecting pairs based on keywords in their commit messages (\eg
\citep{DBLP:conf/sigsoft/LuoHEM14}).

\begin{table}[t]
\small
  \caption{Top Projects with Artifacts}
  \label{tab:top-projects}
  \centering
  \setlength{\tabcolsep}{2pt}
  \begin{tabular}{ lrlr }
    \toprule

    Java & \# Pairs & Python & \# Pairs \\
\midrule
raphw/byte-buddy   & 361 & terasolunaorg/guideline & 171 \\
checkstyle/checkstyle & 184 & scikit-learn/scikit-learn & 151 \\
square/okhttp & 104 & numpy/numpy & 145 \\
HubSpot/Baragon & 94 & python/mypy & 114 \\
tananaev/traccar & 59 & marshallward/f90nml & 65 \\

    \bottomrule
  \end{tabular}
\end{table}


Among all the pairs that we attempted to reproduce, most were not
reproducible. In other words, the log of the original job and the log
produced by \reproducer were different. To gather information about
the causes of unreproducibility, we randomly sampled 100
unreproducible job pairs and manually inspected their 200 logs (two
logs per job pair). For this task, we also examined the corresponding
200 original logs produced by \tvis to compare the differences between
logs and categorize the various sources of unreproducibility.

As shown in ~\cref{tab:unreproducible}, we identified 6 sources of
unreproducibility. From the 200 jobs, around 30\% are unreproducible
due to missing or incompatible dependencies. Another 30\% referenced
stale URLs or experienced network issues. Exceptions from invoking
\tvisbuild when creating build scripts are responsible for
another 20\%. The rest of the jobs are unreproducible due to project-specific
issues, failure to terminate within the time budget,
or permission errors. Interestingly, 6 jobs are actually
reproducible, but since the corresponding failed or passed job is not
reproducible, the entire pair is marked as unreproducible. We have not
included unreproducible pairs in this iteration of \bugswarm, but we think these
could also be potentially useful to researchers interested in
automatically fixing broken builds.

\begin{table}[t]
\small
  \caption{Sources of Unreproducibility}
  \label{tab:unreproducible}
  \centering
  \setlength{\tabcolsep}{4pt}
  \begin{tabular}{ lr }
    \toprule

    Reason                               & \# Pairs \\
    \midrule
    Failed to install dependency   &       59 \\
    URL no longer valid or network issue &       57 \\
    \tvis command issue                  &       38 \\
    Project-specific issue               &       22 \\
    \reproducer did not finish           &       14 \\
    Permission issue                     &        4 \\
    \midrule
    Total                                &      194 \\

    \bottomrule
  \end{tabular}
\end{table}

\begin{table}[t]
\small
  \caption{Diversity of Artifacts}
  \label{tab:diversity}
  \centering
  \setlength{\tabcolsep}{4pt}
  \begin{tabular}{ lrlr }
    \toprule

    Type & \# Artifacts & Type & \# Artifacts \\
\midrule
\cellcolor{light-gray} Language    & \cellcolor{light-gray} &
\cellcolor{light-gray} Longetivity & \cellcolor{light-gray} \\

\MyIndent Java   & 1,827 & \MyIndent 2015; 2016 &   790; 989 \\
\MyIndent Python & 1,264 & \MyIndent 2017; 2018 & 807; 515 \\

\cellcolor{light-gray} Build System   & \cellcolor{light-gray} &
\cellcolor{light-gray} Test Framework & \cellcolor{light-gray} \\

\MyIndent Maven  & 1,675 & \MyIndent JUnit    &   768 \\
\MyIndent Gradle &    86 & \MyIndent unittest &   665 \\
\MyIndent Ant    &    66 & \MyIndent Others   & 1,415 \\

    \bottomrule
  \end{tabular}
\end{table}

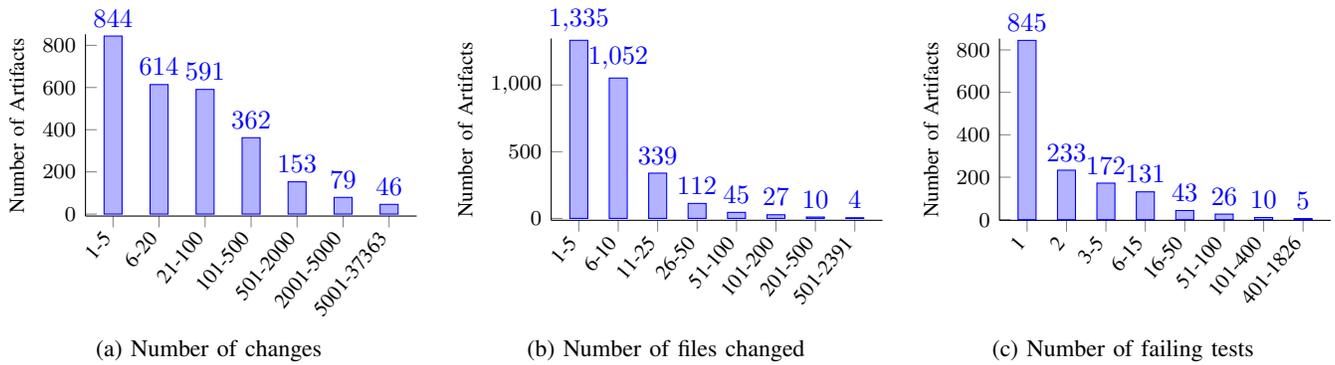
\begin{figure*}[t]
\begin{subfigure}[b]{0.33\textwidth}
  \begin{center}
    \begin{tikzpicture}
      \begin{axis}[
        ybar=-\textwidth/20,
        axis x line*=bottom,
        axis y line*=left,
        height=4cm, 
        ymin=-1,
        width=\textwidth,
        bar width=\textwidth/25,
        ylabel={Number of Artifacts},
        ylabel style={font=\footnotesize},
        y tick label style={font=\footnotesize},
        symbolic x coords = {1-5,6-20,21-100,101-500,501-2000,2001-5000,5001-37363},
        xtick = {1-5,6-20,21-100,101-500,501-2000,2001-5000,5001-37363},
        x tick label style={font=\footnotesize, rotate=50, align=right, anchor=east},
        enlarge y limits=0.01,
        nodes near coords,
        nodes near coords align={vertical}          
        ]
        \addplot coordinates 
        {(1-5,844)
(6-20,614)
(21-100,591)
(101-500,362)
(501-2000,153)
(2001-5000,79)
(5001-37363,46)
}; 
      \end{axis}
    \end{tikzpicture}
  \end{center}
  \vspace{-2ex}
  \caption{Number of changes}   
  \label{fig:diffsize_histogram}
\end{subfigure}
\begin{subfigure}[b]{0.33\textwidth}
  \begin{center}
    \begin{tikzpicture}
      \begin{axis}[
        ybar=-\textwidth/25,
        axis x line*=bottom,
        axis y line*=left,
        height=4cm, 
        ymin=-1,
        width=\textwidth,
        bar width=\textwidth/25,
        ylabel={Number of Artifacts},
        ylabel style={font=\footnotesize},
        y tick label style={font=\footnotesize},
        symbolic x coords = {1-5,6-10,11-25,26-50,51-100,101-200,201-500,501-2391},
        xtick = {1-5,6-10,11-25,26-50,51-100,101-200,201-500,501-2391},
        x tick label style={font=\footnotesize, rotate=50, align=right, anchor=east},
        enlarge y limits=0.01,
        nodes near coords,
        nodes near coords align={vertical}          
        ]
        \addplot coordinates 
	{(1-5,1335)
(6-10,1052)
(11-25,339)
(26-50,112)
(51-100,45)
(101-200,27)
(201-500,10)
(501-2391,4)}; 
      \end{axis}
    \end{tikzpicture}
  \end{center}
  \vspace{-1ex}
  \caption{Number of files changed}   
  \label{fig:diffcount_histogram}
\end{subfigure}
\begin{subfigure}[b]{0.33\textwidth}
  \begin{center}
    \begin{tikzpicture}
      \begin{axis}[
        ybar=-\textwidth/25,
        axis x line*=bottom,
        axis y line*=left,
        height=4cm,
        width=\textwidth,
        bar width=\textwidth/25,
        ylabel={Number of Artifacts},
        ylabel style={font=\footnotesize},
        y tick label style={font=\footnotesize},
        symbolic x coords = {1,2,3-5,6-15,16-50,51-100,101-400,401-1826},
        xtick = {1,2,3-5,6-15,16-50,51-100,101-400,401-1826},
        ytick = {0, 200, 400, 600,800,1000},
        x tick label style={font=\footnotesize, rotate=50, align=right, anchor=east},
        enlarge y limits=0.01,
        nodes near coords,
        nodes near coords align={vertical}          
        ]
        \addplot coordinates 
        {(1,845)
(2,233)
(3-5,172)
(6-15,131)
(16-50,43)
(51-100,26)
(101-400,10)
(401-1826,5)}; 
      \end{axis}
    \end{tikzpicture}
  \end{center}
  \vspace{-1ex}
  \caption{Number of failing tests}
  \label{fig:failedtest_histogram}
\end{subfigure}
\caption{Artifact Characteristics}
\label{fig:characteristics}
\end{figure*}

\begin{mdframed}[backgroundcolor=gray!20]
  \small \textbf{RQ2}: Reproducing fail-pass pairs is indeed
  challenging with a 5.56\% success rate. Based on the manual
  inspection of 100 unreproducible artifacts, we identified 6 main
  reasons for unreproducibility listed in \cref{tab:unreproducible}.
\end{mdframed}

\subsection{General Characteristics of \bugswarm Artifacts}

We have aggregated statistics on various artifact
characteristics. \cref{fig:diffsize_histogram} shows the number of
artifacts with a given number of changes (additions or deletions).
Inserting or removing a line counts as one change. Modifying an
existing line counts as \textit{two} changes (an addition and a
deletion). Commits with zero changes are possible but rare and are not
included in \cref{fig:diffsize_histogram}. We report the number of
changes of the fixed version with respect to the failing version of
the code, \eg 31\% (844) of the artifacts have at most 5 changes and
54\% (1,458) have at most 20. \cref{fig:diffcount_histogram} shows the
number of projects with a given number of files changed, \eg 46\%
(1,335) of the artifacts have at most 5 changed
files. \cref{fig:failedtest_histogram} shows the artifacts with a
given number of failed tests.

\begin{figure*}
\begin{subfigure}[b]{0.33\textwidth}
  \begin{center}
    \begin{tikzpicture}
        \begin{axis}[xbar,
        xbar=-\textwidth/65,
        axis x line*=bottom,
        axis y line*=left,
        xmin=-1,xmax=100,
        height=5cm, 
        width=\textwidth/1.5,
        bar width=\textwidth/65,
        xlabel={Number of Artifacts},
        x label style={font=\footnotesize},
        x tick label style={font=\footnotesize},
        xtick={0,50,100},
        symbolic y coords = {Resource leak,Casting error,Visibility error,Identifier error,Dependency error,%
        Configuration error,NullPointerException,Assertion error,%
        Test error,Logic error},
        ytick = {Resource leak,Casting error,Visibility error,Identifier error,Dependency error,%
        Configuration error,NullPointerException,Assertion error,%
        Test error,Logic error},
        yticklabels = {Resource Leak,Casting Error,Visibility Error,Identifier Error,Dependency Error,%
        Configuration Error,NullPointerException,Assertion Error,%
        Test Error,Logic Error},
        y tick label style={font=\footnotesize, rotate=0, align=right, anchor=east},
        nodes near coords,
        nodes near coords align={horizontal}          
        ]
        \addplot coordinates {(1,Resource leak)(2,Casting error)(7,Visibility error)(9,Identifier error)(14,Dependency error)(28,Configuration error)(32,NullPointerException)(38,Assertion error)(50,Test error)(94,Logic error)};
     \end{axis}
  \end{tikzpicture} 
  \end{center}
  \caption{Manual Classification of Java Bugs}
  \label{fig:manual-classification}
\end{subfigure}
\begin{subfigure}[b]{0.33\textwidth}
  \begin{center}
    \begin{tikzpicture}
      \begin{axis}[xbar,
        xbar=-\textwidth/65,
        axis x line*=bottom,
        axis y line*=left,
        xmin=-1,xmax=400,
        height=5cm, 
        width=\textwidth/1.5,
        bar width=\textwidth/65,
        xlabel={Number of Artifacts},
        x label style={font=\footnotesize},
        xtick={0,200,400},
        x tick label style={font=\footnotesize},
        symbolic y coords = {SAXParseException,FileNotFoundException,%
         SocketException,IllegalArgumentException,IOException,ClassNotFoundException,%
        RuntimeException,IllegalStateException,NullPointerException,AssertionError},
        ytick={SAXParseException,FileNotFoundException,%
         SocketException,IllegalArgumentException,IOException,ClassNotFoundException,%
        RuntimeException,IllegalStateException,NullPointerException,AssertionError},
        yticklabels = {SAXParseExc.,FileNotFoundExc.,SocketExc.,IllegalArgumentExc.,%
        IOExc.,ClassNotFoundExc.,RuntimeExc.,IllegalStateExc.,NullPointerExc.,AssertionErr.},
        y tick label style={font=\footnotesize, rotate=0, align=right, anchor=east},
        nodes near coords,
        nodes near coords align={horizontal}          
        ]
        \addplot coordinates {(298,AssertionError)
(252,NullPointerException)
(159,IllegalStateException)
(150,RuntimeException)
(136,ClassNotFoundException)
(85,IOException)
(76,IllegalArgumentException)
(50,SocketException)
(46,FileNotFoundException)
(38,SAXParseException)};
      \end{axis}
    \end{tikzpicture}
    \end{center}
  \caption{Most Frequent Java Exceptions}
  \label{fig:java_error_10}
\end{subfigure}
\begin{subfigure}[b]{0.33\textwidth}
  \begin{center}
    \begin{tikzpicture}
      \begin{axis}[xbar,
        xbar=-\textwidth/65,
        axis x line*=bottom,
        axis y line*=left,
        xmin=-1,xmax=500,
        height=5cm, 
        width=\textwidth/1.5,
        bar width=\textwidth/65,
        xlabel={Number of Artifacts},
        x label style={font=\footnotesize},
        xtick={0,250,500},
        x tick label style={font=\footnotesize},
         symbolic y coords = {IOError,RuntimeError,FileNotFoundError,SyntaxError,%
        NameError,ImportError,TypeError,ValueError,AttributeError,AssertionError},
        ytick={IOError,RuntimeError,FileNotFoundError,SyntaxError,%
        NameError,ImportError,TypeError,ValueError,AttributeError,AssertionError},
        yticklabels = {IOErr.,RuntimeErr.,SyntaxErr.,FileNotFoundErr.,NameErr.,ImportErr.,TypeErr.,%
        ValueErr.,AttributeErr.,AssertionErr.},
        y tick label style={font=\footnotesize, rotate=0, align=right, anchor=east},
        nodes near coords,
        nodes near coords align={horizontal}          
        ]
        \addplot coordinates {(320,AssertionError)
(104,AttributeError)
(97,ValueError)
(86,TypeError)
(64,ImportError)
(31,NameError)
(29,SyntaxError)
(28,FileNotFoundError)
(26,RuntimeError)
(24,IOError)};
      \end{axis}
    \end{tikzpicture}
  \end{center}     
  \caption{Most Frequent Python Errors}
  \label{fig:py_errors_10}
\end{subfigure}
\caption{Artifact Classification}
\label{fig:classification}
\end{figure*}
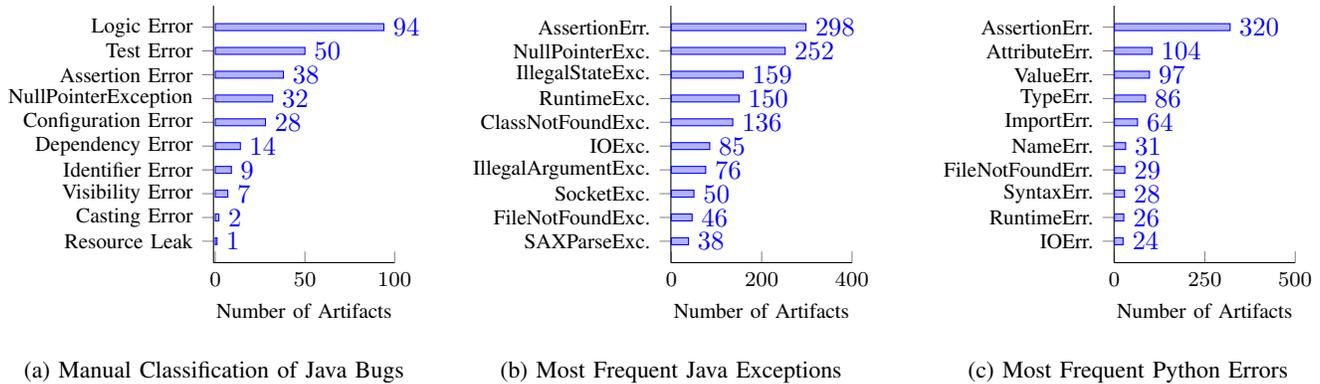

We find that our artifacts are diverse in several aspects: language,
build system, test framework, and longevity.~\cref{tab:diversity}
shows the number of reproducible and flaky artifacts for each of these
categories. The current dataset has over a thousand artifacts for Java
and Python with a wide range of build systems and testing frameworks
being used. From these, the most common build system is Maven with
1,675 artifacts, and the most common testing framework is JUnit with
768.  We plan to add support for other languages such as JavaScript
and C++ in the near future, which will increase the number of build
systems and testing frameworks being used.

Our artifacts represent a variety of software bugs given the diverse
set of projects mined. To better understand the types of bugs in
\bugswarm, we conduct a manual classification of \mnlclasssize
randomly sampled Maven-based Java artifacts, first described in
\citep{DBLP:conf/sigsoft/Tomassi18}. The top \nummnlclassctgrs
classification categories are shown in
\cref{fig:manual-classification}. The classification is not
one-to-one; an artifact may fall under multiple categories depending
on the bug. To correctly classify an artifact, we examine the source
code, diff, commit message, and \tvis log. We find that the
largest category is logic errors. Examples of logic errors include
off-by-one errors and incorrect logical operations.

We also conduct an automatic higher-level classification of artifacts
based on the encountered exceptions or runtime errors. We analyze the
build logs and search for the names of Java exceptions and Python
runtime errors. \cref{fig:java_error_10,fig:py_errors_10} show the 10
exceptions/errors for which \bugswarm has the most artifacts. For
example, 252 Java artifacts fail with a
NullPointerException. An example is shown in
\cref{fig:yamcs-yamcs-186324159}.

\begin{figure}
\begin{lstlisting}
 protected void loadCommandVerificationSheet(
     SpaceSystem spaceSystem, String sheetName) {
   Sheet sheet = switchToSheet(sheetName, false);
+  if (sheet == null) return;
   int i = 1;
   while (i < sheet.getRows()) {
     // search for a new command definition
     ...
   }
 }
   
\end{lstlisting}
  \vspace{-2ex}
  \caption{Example of NullPointerException bug and its fix.}
  \label{fig:yamcs-yamcs-186324159}
\end{figure}


\label{cobertura}

Using the \bugswarm framework presented in
\cref{sec:artifact-execution}, we successfully ran the code coverage
tool Cobertura \citep{cobertura} and two static analyzers\textemdash
Google's ErrorProne \citep{errorprone} and SpotBugs
\citep{spotbugs}\textemdash on the 320 randomly selected artifacts
used in the manual classification \citep{DBLP:conf/sigsoft/Tomassi18},
with minimal effort.

\begin{mdframed}[backgroundcolor=gray!20]
  \small \textbf{RQ3}: We investigated various characteristics of
  artifacts, such as the distribution in the size of the diff,
  location of the diff, and number of failing tests
  (\cref{fig:characteristics}). We also examined the reason for
  failure (\cref{fig:classification}). For example, 844 artifacts have
  between 1 and 5 changes, 1,335 artifacts modify a single file, 845
  artifacts have 1 failing test, and the top reason for a build
  failure is an AssertionError.
\end{mdframed}

\subsection{Performance}

\pairminer and \reproducer can be run in the cloud in
parallel.
The \bugswarm infrastructure provides support to run these as batch
tasks on Microsoft Azure \citep{azure}. Running time of \pairminer
depends on the number of failed jobs to examine, taking between a few
minutes to a few hours. Reproduction time varies per project as it
depends on the project's build time and the number of tests
run. Mining and reproducing the pairs reported in this paper required
about 60,000 hours of compute time in Azure. We will continue our
effort to mine and reproduce pairs in additional
projects.


\section{Limitations and Future Work}
\label{sec:limitations}
\pairminer searches for two consecutive failed and passed builds first
then looks for failed and passed job pairs within these two builds.
However, failed and passed job pairs can occur between two consecutive
failed builds because a build marked as failed requires only one
unsuccessful job.  In addition, the fail-pass pattern does not
guarantee that the difference between the two commits is actually a
fix for the failure; the supposed fix could simply delete or revert
the buggy code or disable any failing tests. Using only the pattern,
\pairminer would also fail to identify a fix for a failure if the fix
is committed along with the test cases that expose the fail
point. Finally, the fix may not be minimal. We plan to address some of
these challenges in the future. In particular, we would like to
explore other mining approaches that involve new patterns as well as
bug reports. Note that \reproducer is already capable of reproducing
any pair of commits that triggered \tvis builds, regardless of how
these commits are gathered.

Reproducible artifacts may still break later on due to stale URLs,
among other reasons. To keep \bugswarm up to date, we periodically
test artifacts. We are currently exploring ways to make the artifacts
more robust. In the future, we would like to crowdsource the
maintainability of \bugswarm.

Thus far, our mining has been ``blind.'' However, it is possible to
extend our mining tools to find pairs with specific characteristics
(\eg pairs that have at most 5 changes and a single failed test caused
by a NullPointerException). Such guided mining will allow \bugswarm to
grow in directions of interest to the research community. Finally, we
plan to extend \bugswarm to continuously monitor \tvis events for
real-time mining and reproducing of new artifacts.


\section{Related Work}
\label{sec:related}
Some other defect repositories aim to provide experimental benchmarks
for
defect location and repair. On the whole, these repositories do not
exploit CI and virtualization mechanisms; they generally pre-date the
widespread adoption of these techniques.  They do not achieve the same
\textit{scale}, \textit{diversity}, and \textit{currency} and are not
as \textit{durably reproducible}.

The Siemens test suite \citep{DBLP:conf/icse/HutchinsFGO94} (7 small C
programs and about 130 manually seeded bugs) is among the
earliest. BugBench \citep{Lu05bugbench:benchmarks} is one of the
earliest datasets of real-world bugs. BugBench is limited in scale and
diversity, consisting of 7 memory and concurrency bugs found across 10
C/C++ OS projects. Each buggy program version includes failing
tests. BegBunch \citep{1555866} contains two suites to measure the
accuracy and scalability of bug detection tools for C. iBugs
\citep{DBLP:conf/kbse/DallmeierZ07} is a dataset drawn from the 5-year
history of the AspectJ compiler with 369 faulty versions of the
project. iBugs provides metadata such as number of methods and classes
involved in the bug fix. Unlike \bugswarm, the above datasets were
manually constructed. Metadata such as that included in iBugs could be
built from \bugswarm artifacts with additional effort.

The Software-artifact Infrastructure Repository (SIR)
\citep{DBLP:journals/ese/DoER05} comprises source code, tests, and
defects from OS projects along with needed infrastructure (\eg
automated build and test scripts). Currently, SIR consists of 85
projects in four languages, of which 64 (15 C, 1 C\#, 1 C++, and 47
Java) include fault data: real ones; seeded ones; and a combination
of real, seeded, and mutated. A project may contain multiple versions,
and each version may contain multiple faults, with a total of 680
bugs. SIR provides a useful amount of scale and diversity while archiving
sufficient tooling for durable reproducibility. However, since it
pre-dates CI and \dockr, each defect datum therein is manually
assembled. Thus, SIR is difficult to scale up further and requires
substantial effort to keep current. \bugswarm already has
\numJobPairArtifacts reproducible defects; the automated mining of CI
and \dockr image artifacts lowers the cost of keeping the
dataset growing.

\textsc{ManyBugs} \citep{DBLP:journals/tse/GouesHSBDFW15} is a
benchmark for program repair with 185 defects and fixes from 9 large C
projects. Each defect and fix includes tests and is manually
categorized. To facilitate the reproduction of these defects,
\textsc{ManyBugs} provides virtual machine images (recently extended
to use \dockr \citep{DBLP:conf/icse/TimperleySG18}). Unlike \bugswarm,
mining and reproducing bugs requires significant manual effort, and
thus \textsc{ManyBugs} is not as easy to extend. On the other hand,
\textsc{ManyBugs} provides a detailed bug categorization that can be
useful for experiments, and its artifacts are collected from C
programs, a programming language that \bugswarm does not currently
support.

Defects4J \citep{DBLP:conf/issta/JustJE14}\citep{defects4j} is a
dataset of 395 real, reproducible bugs from 6 large Java
projects. Defects4J provides manually constructed scripts for each
project's build and test; the entire setup relies on a functioning
JVM. Defects4J provides an interface for common tasks and provides
support for a number of tools. The Bugs.jar
\citep{DBLP:conf/msr/SahaLLYP18} dataset contains 1,158 real,
reproducible Java bugs collected from 8 Apache projects by identifying
commit messages that reference bug reports. Bugs.jar artifacts are
stored on \git branches. By contrast, \bugswarm relies on virtualized,
\dockr-packaged build and test environments, automatically harvested
from the cross-platform \tvis archives; thus it is neither limited to
Java nor does it require manual assembly of build and test tools. In
addition to the test fail-pass pairs, we include build failures and
even flaky tests. The above allows \bugswarm to achieve greater scale,
diversity, and currency.

\citet{urli2018design} describe an approach to mining builds that fail
tests from \tvis. This work can only handle Maven-based Java builds;
these are reproduced directly, without \dockr. Their dataset includes
3,552 Maven Java builds for the purpose of automatic
repair. \citet{bears} develop \textsc{Bears}, which mines Maven-based
Java \ghub projects that use \tvis.  \textsc{Bears} attempts to
reproduce every mined build in the same environment, which does not
account for the developer-tailored \tvisyml file, whereas \bugswarm
leverages \dockr images to match each job's original runtime
environment.  Compared to \bugswarm, \textsc{Bears} has a similar
reproduction success rate of 7\% (856 builds).  \textsc{Bears} pushes
artifacts to \git branches, instead of providing them as \dockr
images, and relies on Maven for building and testing, so new
infrastructure must be implemented to include artifacts from other
build systems.

Our \dockr-based approach allows other languages and build systems,
and reflects our designed-in pursuit of greater diversity and
reproducibility. Note that the \bugswarm toolset supports the creation
of fully reproducible packages for any pair of commits for which the
\tvis builds are archived. There are over 900K projects in \ghub that
use \tvis\citep{travis-ci}, so our toolkit enables the creation of
datasets and ensuing experiments at a scale substantially larger than
previous datasets allow.


\section{Conclusions}
\label{sec:conclusions}
This paper described \bugswarm, an approach that leverages CI to mine
and reproduce fail-pass pairs of realistic failures and fixes in Java
and Python OSS. We have already gathered \numJobPairArtifacts such
pairs.  We described several exciting future directions to further
grow and improve the dataset. We hope \bugswarm will minimize effort
duplication in reproducing bugs from OSS and open new research
opportunities to evaluate software tools and conduct large-scale
software studies.

\section*{Acknowledgments}

We thank Christian Bird, James A. Jones, Claire Le Goues, Nachi
Nagappan, Denys Poshyvanyk, Westley Weimer, and Tao Xie for early
feedback on this work. We also thank Saquiba Tariq, Pallavi Kudigrama,
and Bohan Xiao for their contributions to improve \bugswarm and Aditya
Thakur for feedback on drafts of this paper. This work was supported
by NSF grant CNS-1629976 and a Microsoft Azure Award.

\balance

\def\bibfont{\footnotesize}
\bibliography{dblp,local}

\end{document}